%% file: arxiv.tex
\documentclass{easychair}
\pdfoutput=1

\usepackage{color}
\usepackage{graphicx}
\usepackage{amssymb,amsmath}
\usepackage{algorithm}
\usepackage[noend]{algpseudocode}
\makeatletter
\let\OldStatex\Statex
\renewcommand{\Statex}[1][3]{%
  \setlength\@tempdima{\algorithmicindent}%
  \OldStatex\hskip\dimexpr#1\@tempdima\relax}
\makeatother
\usepackage{xspace}

\newcommand{\toolname}{\textsc MapleCrypt\xspace}
\newcommand{\maplesat}{\textsc MapleSAT\xspace}

\begin{document}

\title{Adaptive Restart and CEGAR-based Solver for Inverting
  Cryptographic Hash Functions}

\author{Saeed Nejati, Jia Hui Liang, Vijay Ganesh,\\
  Catherine Gebotys \and Krzysztof Czarnecki}

\institute{University of Waterloo, Waterloo, ON, Canada\\
}

\authorrunning{Nejati, Liang, Ganesh, Gebotys and Czarnecki}
\titlerunning{Adaptive Restart and CEGAR-based Solver}

\maketitle


\input{abstract.tex}


\input{sec.intro.tex}


\input{sec.background.tex}


\input{sec.proposed.tex}


\input{sec.results.tex}


\input{sec.related.tex}


\input{sec.conclusion.tex}




%
\bibliographystyle{plain}
\bibliography{arxiv}

\end{document}

%% file: abstract.tex
\begin{abstract}
  SAT solvers are increasingly being used for cryptanalysis of hash
  functions and symmetric encryption schemes. Inspired by this trend,
  we present \toolname which is a SAT solver-based cryptanalysis tool
  for inverting hash functions. We reduce the {\it hash function
    inversion problem for fixed targets} into the satisfiability
  problem for Boolean logic, and use \toolname to construct preimages
  for these targets. \toolname has two key features, namely, a
  multi-armed bandit based adaptive restart (MABR) policy and a
  counterexample-guided abstraction refinement (CEGAR) technique. The
  MABR technique uses reinforcement learning to adaptively choose
  between different restart policies during the run of the solver. The
  CEGAR technique abstracts away certain steps of the input hash
  function, replacing them with the identity function, and verifies
  whether the solution constructed by \toolname indeed hashes to the
  previously fixed targets. If it is determined that the solution
  produced is spurious, the abstraction is refined until a correct
  inversion to the input hash target is produced. We show that the
  resultant system is faster for inverting the SHA-1 hash
  function than state-of-the-art inversion tools.
\end{abstract}

%% file: sec.intro.tex
\section{Introduction}
\label{sec:intro}

Over the last 15 years we have seen a dramatic improvement in the efficiency of
conflict-driven clause-learning (CDCL) SAT
solvers~\cite{marques1999grasp,moskewicz2001chaff,audemard2009glucose,biere2010lingeling}
over industrial instances generated from a large variety of applications such
as verification, testing, security, and
AI~\cite{cadar2008exe,biere2003bounded,rintanen2009planning}.  Inspired by this
success many researchers have proposed the use of SAT solvers for cryptanalysis
of hash functions and symmetric encryption
schemes~\cite{mironov2006applications}. The use of SAT solvers in this context
holds great promise as can be seen based on their success to-date in automating
many aspects of analysis of cryptographic primitives~\cite{aarontomb}. SAT
solvers are increasingly an important tool in the toolbox of the practical
cryptanalyst and designer of hash functions and encryption schemes. Examples of
the use of SAT solvers in cryptanalysis include tools aimed at the search for
cryptographic keys in 1999 \cite{massacci1999using}, logical cryptanalysis as a
SAT problem in 2000 \cite{massacci2000logical}, encoding modular root finding
as a SAT problem in 2003 \cite{fiorini2003fake} and logical analysis of hash
functions in 2005 \cite{jovanovic2005logical}.  Most of these approaches used a
direct encoding of the said problems into a satisfiability problem and used SAT
solvers as a blackbox.

In this paper, we propose a set of techniques and an implementation, we call
\toolname, that dramatically improve upon the state-of-the-art in solving the
{\it cryptographic hash function inversion problem}. The problem of inverting
hash function is of great importance to cryptographers and security
researchers, given that many security protocols and primitives rely on these
functions being {\it hard-to-invert}. Informally, the problem is ``given a
specific hash value (or target) $H$ find an input to the hash function that
hashes to $H$''. We focus on the inversion problem, as opposed to the more
well-studied collision problem. The value of this research is not only the fact
that cryptanalysis is an increasingly important area of application for SAT
solvers, but also that instances generated from cryptographic applications tend
to be significantly harder for solvers than typical industrial instances, and
hence are a very good benchmark for solver research.

{\bf Summary of Contributions.} We focus on the SHA-1 cryptographic hash
function in this paper, and make the following contributions:

{\bf 1.} We present a counter-example guided
abstraction-refinement~\cite{clarke2000counterexample} (CEGAR) based technique,
wherein certain steps of the hash function under analysis are abstracted away
and replaced by the identity function. The inversion problem for the resultant
abstracted hash function is often much easier for solvers, and we find that we
do not have to do too many steps of refinement. An insight from this experiment
is that certain steps of the SHA-1 hash function do not have sufficient levels
of diffusion, a key property of hash functions that makes them difficult to
invert.

{\bf 2.} In addition to the above-mentioned CEGAR technique, we present a
multi-armed bandit~\cite{sutton1998introduction} based adaptive restart policy.
The idea is that a reinforcement learning technique is used to select among a
set of restart policies in an online fashion during the run of the solver. This
method is of general value, beyond cryptanalysis. The result of combining these
two techniques is a tool, we call \toolname, that is around two times faster on
the hash function inversion problem than most state-of-the-art SAT solvers.
More importantly, with a time limit of 72 hours, \toolname can invert a 23-step
reduced version of SHA-1 consistently, whereas other tools we compared against
can do so only occasionally.

{\bf 3.} We perform extensive evaluation of \toolname on the SHA-1 inversion
problem, and compare against CryptoMiniSat, Lingeling, MiniSAT, Minisat-BLBD.
In particular, \toolname is competitive against the best tools out there for
inverting SHA-1.

%% file: sec.background.tex
\section{Background on Cryptographic Hash Functions}
\label{sec:background}
In this section we provide a brief background on cryptographic hash
functions, esp. SHA-1. We refer the reader to ~\cite{fips2011180} for
a more detailed overview of hash functions. A hash function maps an
arbitrary length input string to a fixed length output string (e.g.,
160 bits in the case of SHA-1).  There are three main properties that
are desired for a cryptographic hash function
\cite{lai1993hash}. Informally, they are:
\begin{itemize}
  \item \textit{Preimage Resistance}: Given a hash value $H$, it
    should be computationally infeasible to find a message $M$, where
    $H = hash(M)$.

  \item \textit{Second Preimage Resistance}: Given a message $M_1$,
    it should be computationally infeasible to find another message
    $M_2$, where $hash(M_1) = hash(M_2)$ and $M_1 \neq M_2$. 

  \item \textit{Collision Resistance}: It should be computationally
    infeasible to find a pair of messages $M_1$ and $M_2$, where
    $hash(M_1) = hash(M_2)$ and $M_1 \neq M_2$. (There is a subtle
    difference between second pre-image resistance and collision
    resistance, in that the message $M_1$ is not fixed in the case of
    collision resistance).
\end{itemize}

Preimage resistance implies that the hash function should be hard to
invert. The terms preimage attack and inversion attack are used interchangeably.
Standard cryptographic hash functions at their core have a compression
function, which can essentially be seen as repeated application of a
step function on its input bits for a fixed number of steps.
The compression function takes as input a fixed length input
and outputs a fixed length (with smaller length) output. For making a
collision resistant compression function, one method is to use a block
cipher and apply the Davis-Meyer method. Feistel ladder which is
widely used in hash functions like MD5 and SHA-1 (and also block
ciphers like DES), is an implementation of this method, where the key
is a message word. If the key is known, each step is easily
reversible. For making a hash function able to accept arbitrary long
messages as input, one can use Merkle-Damgard structure
\cite{merkle1989one}, where it is shown that if one block is collision
resistant, the whole structure would be collision resistant.

\subsection{SHA-1}
\label{sec:sha1}
SHA-1 (Secure Hash Algorithm), was designed by NSA, and adopted as a
standard in 1995 \cite{fips2011180}, and is still widely used in many
applications. SHA-1 consists of iterative application of a so-called
compression function which takes a 160-bit chaining value and
transforms it into the next chaining value using a 512-bit message
block. The current chaining value would be added to the output of
compression function to make the next chaining value: $CV_{i+1}=Comp(CV_i, M_i)+CV_i$,
and the $CV_0$ is set to a fixed initialization vector.
Before starting the process, the message is padded with a
single bit '1' followed by a set of '0's and original message length
as a 64-bit value. The length of padded version is a multiple of 512.
The message is broken down to blocks of 512 bits. Each compression
function breaks down the input message block into sixteen 32-bit words.
Then it will go through a message expansion phase, which extends
sixteen words to eighty words, using the following formulation
(consider that $W_i$ for $0 \leq i \leq 15$ refers to input message words):

\begin{equation}
W_i = (W_{i-3} \oplus W_{i-8} \oplus W_{i-14} \oplus W_{i-16})\lll 1 \quad (16 \leq i \leq 79)
\end{equation}
where '$\lll 1$' is left rotation by one position.

There are five 32-bit words (namely, $a, b, c, d, e$) as the
intermediate variables.  In each step a function $F_t$ is applied to
three of these words, and it changes every 20 steps:

\begin{equation} \label{eq:nonlinear}
F_t(b, c, d) = \left\lbrace
  \begin{array}{l l}
    Ch(b, c, d) = (b \land c) \oplus (\neg b \land d) & \quad 0 \leq t \leq 19 \\
    Parity(b, c, d) = b \oplus c \oplus d & \quad 20 \leq t \leq 39 \\
    Maj(b, c, d) = (b \land c) \oplus (b \land d) \oplus (c \land d) & \quad 40 \leq t \leq 59 \\
    Parity(b, c, d) = b \oplus c \oplus d & \quad 60 \leq t \leq 79 \\
  \end{array} \right.
\end{equation}

The step process would be like:

\begin{multline} \label{eq:step}
(a_{t+1}, b_{t+1}, c_{t+1}, d_{t+1}, e_{t+1}) \leftarrow \\
(F_t(b_t, c_t, d_t) \boxplus e_t \boxplus (a_t \lll 5) \boxplus W_t \boxplus K_t, a_t, b_t \lll 30, c_t, d_t)
\end{multline}
where $\lll$ is left rotation, $\boxplus$ is modulo-$2^{32}$ addition
and $K_t$ is the round constant. This is repeated for next 512-bit
message block in the Merkel-Damagard chain.

\noindent{\bf Step-Reduced Version:} Usually inverting or finding
collision for full version of a hash function is very hard. Thus,
cryptanalysts work on a relaxed version of those functions like
step-reduced versions, which means the function under attack is the
same, except that the number of steps is reduced.

%% file: sec.proposed.tex
\section{Architecture of \toolname}
\label{sec:proposed}
We seek to perform a preimage or inversion attack on a step-reduced version of
SHA-1, with one block input (512 bits).  Although the current work is focused
on SHA-1, our approach is applicable to other iterative hash functions. The two
main contributions in our design are adaptive restart and a CEGAR-based
approach. The adaptive restart is not directly dealing with the structure of
the function and therefore could be used in solving other SAT instances. The
CEGAR approach is abstracting and refining step functions. Thus it could be
mounted on the other hash functions that have a repeated use of a step function
(e.g. MD4, MD5, SHA-2). 

\subsection{SAT Encoding} The encoding we use in this paper is based on the
given in \cite{nossum2012sat}. Most of the operations are encoded using
Tseitin transformation, but some operations are described using high level
relations.  There is a 5-operand addition in each step of SHA-1 (refer to
equation \eqref{eq:step}).  The main contribution of the encoding in
\cite{nossum2012sat} is the encoding of this multi-operand addition (instead
of encoding of multiple two-operand additions). Current SHA-1 instances in
SAT competition are generated using this tool \cite{nossum2013instance}.  We
have made minor modifications, namely to the encoding of round-dependent
logical function ($F_t$ in equation \eqref{eq:nonlinear}), replacing XOR
operations with inclusive-OR to simplify the corresponding clauses.

\subsection{CEGAR Loop Design}  The SHA-1 SAT instances of up to 20 steps are
very easy to solve (less than a second using most modern solvers), but the
level of difficulty rapidly rises from 20 steps to 23 steps (needs 2 to 3 days
to solve).  To the best of our knowledge, preimage for more than 23 steps
cannot be constructed in a reasonable amount of time even with the latest
techniques and hardware
\cite{nossum2012sat,legendre2014logical,legendre2012encoding}.

For the instances of more than 20 steps (e.g. 22 steps), we abstract
away initial step functions and keep the last 20 steps intact
(abstracting first 2 steps), but we do not abstract away the message
words and the message expansion relations.  We solve the simplified
instance, and find a solution for the message words. However, the
resultant solution may be spurious, i.e., may not actually hold for
the specific hash target.  In order to verify the solution, we run the
hash function in the forward direction and check the result with the
target, and record values of intermediate variables throughout the
hashing.  If the computed hash does not match the target (i.e., the
solution produced by the solver is spurious), we refine back those
parts of abstracted steps of the hash function that are unsatisfiable
under the spurious solution. Finally we also add a subset (all except
last 8 steps) of intermediate values (computed during the forward run
of the hash function on spurious solutions) as blocking clauses.

The intuition behind this procedure is that, first of all, 20 steps
are very easy to solve, and it is the highest number of steps that we
are better off solving directly, rather than using an abstraction.
Secondly, the first few intermediate variables have the most degree of
freedom when searching for a preimage or collision. Lastly, blocking a
subset of intermediate values, although might block some legitimate
solutions, but also blocks many spurious solutions. We can divide our
main procedure into two main functions, listed in Algorithm~\ref{alg:gac}.

\begin{algorithm}[t]
	\caption{Finding Preimage using a CEGAR loop}
	\label{alg:gac}
	\begin{algorithmic}[1]
        \Require $W$: 512-bit found preimage, $H$: Hash target, $nsteps$: number of hash steps
        \Ensure \textit{true} if $W$ is a valid preimage, \textit{false} otherwise
		\Function{Check}{$W, H, nsteps$}
		\State ($H'$, interValues) $\gets$ SHA-1($W$, nsteps)
		\If{$H = H'$}
		\State \Return true
		\Else
		\State CCDB.add(interValues)
		\State \Return false
		\EndIf
		\EndFunction \\

        \Require $H$: Hash target, $nsteps$: number of hash steps
        \Ensure $W$: 512-bit preimage of $H$
        \Function{FindPreimage}{$H, nsteps$}
		\State InstanceSteps~$\gets$ AbsInstGen(nsteps) \Comment{Abstracted set of step functions}
		\State InstanceW $\gets$ MsgInstGen(nsteps)     \Comment{All of input and expanded message words}
		\While{true}
		\State $W[0..15]$ $\gets$ SATSolver(InstanceSteps, InstanceW, CCDB)
		\If{Check(W, H, nsteps) = true}
		\State \Return W
		\EndIf
		\State InstanceSteps $\gets$ Refine(InstanceSteps, W, nsteps)
		\EndWhile
		\EndFunction
	\end{algorithmic}
\end{algorithm}

In the listing of algorithm~\ref{alg:gac}, \textit{interValues} refers
to the collection of intermediate variables across working steps that
will be negated and added as a conflict clause to a CCDB (conflict
clause database).  The \textit{Refine} function evaluates the clauses
of the original formula with the found solution and checks which
variables are in the UNSAT clauses and add them back to the current
abstracted instance.

\subsection{Multi-Armed Bandit Restart}  Many restart policies have been
proposed in the SAT
literature~\cite{biere2008adaptive,audemard2012refining,luby1993optimal,biere2008picosat},
in particular we focus on the uniform, linear, luby, and geometric
restart policies~\cite{biere2015evaluating}. For a given preimage
attack instance, we can not know a priori which of the 4 restart
policy will perform the best. To compensate for this, we use
multi-armed bandits (MAB)~\cite{sutton1998introduction}, a special
case of reinforcement learning, to switch between the 4 policies
dynamically during the run of the solver. We chose to use discounted
UCB algorithm~\cite{garivier2011} from MAB literature, as it accounts
for the nonstationary environment of the CDCL solver, in particular
changes in the learnt clause database over time. Discounted UCB has 4
actions to choose from corresponding to the uniform, linear, luby, and
geometric restart policies. Once the action is selected, the solver
will proceed to perform the CDCL backtracking search until the chosen
restart policy decides to restart. The algorithm computes the average
LBD (Literals Block Distance~\cite{audemard2009predicting}) of the
learnt clauses generated since the action was selected, and the
reciprocal of the average is the reward given to the selected action.
Intuitively, a restart policy which generates small LBDs will receive
larger rewards and UCB will increase the probability of selecting that
restart policy in the future. Over time, this will bias UCB towards
restart policies that generate small LBDs.

%% file: sec.results.tex
\section{Experimental Results}
\label{sec:results}
\subsection{Experimental Setup}  Our baseline benchmark consists of instances
of preimage of step-reduced SHA-1, from 21 to 23 steps. Instances for less than
21 steps were trivial for every solver we tried.  For each step we generated 25
random targets and encoded them as fixed value for the hash output.  All jobs
were run on AMD opteron CPUs at 2.2 GHz and 8GB RAM.  The timeout for solving a
single instance was 72 hours, with 4GB of memory allocated for each process. We
used 5 SAT solvers, CryptoMiniSat-4.5.3 \cite{cryptominisat},
Minisat\_BLBD~\cite{chen2014bit}, Lingeling-ayv~\cite{lingeling},
Minisat-2.2~\cite{minisat} and \maplesat~\cite{liang2016maplesat}.  We also
tried 4 SMT solvers to take advantage of their BitVector theory solvers. We
used, STP-2.0/CryptoMinisat4, Boolector-2.0.7/Lingeling, CVC4-1.5, Z3-4.4.
However, all of these SMT solvers performed surprisingly poorly. Hence, we did
not include them in our comparisons.

\subsection{CEGAR and MABR} 
\begin{figure}[t]
    \centering
    \includegraphics[width=0.8\textwidth]{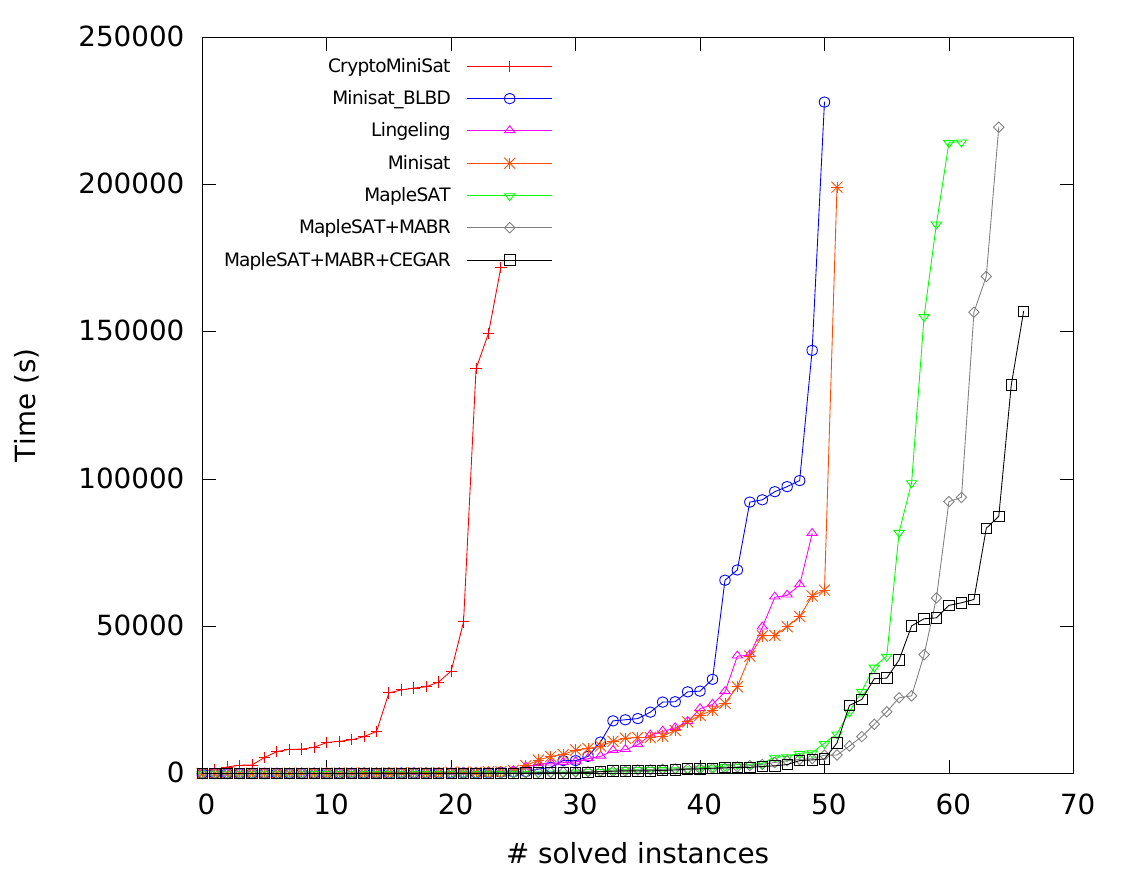}
    \caption{The performance of various SAT solvers against \maplesat
      with adaptive MAB-restart and CEGAR.}
    \label{fig:cegar}
\end{figure}

State-of-the-art results for automated and
practical preimage of SHA-1 that construct a result rather than presenting an
upper bound for attack complexity, propose a SAT encoding of the preimage
attack and solve it using modern SAT solvers.  Therefore we picked the best
existing encoding method for SHA-1~\cite{nossum2012sat} and applied our solving
techniques on them. We are comparing our runtimes with other SAT solvers, given
the same instances. Figure~\ref{fig:cegar} shows the cactus plot of solving
times where each data point shows how many instances could be
solved in the corresponding time. Curves more toward bottom are faster
and more toward right are solving more instances. It can be seen that \maplesat
with MAB restart dominates in terms of runtime and number of instances solved,
and after employing the CEGAR technique, we are able to solve faster and more
instances. MapleCrypt is the CEGAR architecture that uses MapleSAT+MABR as
backend solver.

\subsection{Partial Preimage} This is the kind of attack where the attacker
knows some bits of the input message and wants to find out the rest. Our
experiments on SHA-1 show that knowing parts of the message does not
necessarily make the problem easier, as it might force the solver to find a
specific input that matches those bits and reduces the possibilities. Our
results mostly confirm the observations on hardness of partial preimage of
SHA-1 in \cite{nossum2013instance}.  We could invert up to 27 steps of SHA-1
when having 40 bits of the input message unknown, which matches the best
results known in this setting for SHA-1 \cite{morawiecki2013sat}.



%% file: sec.related.tex
\section{Related Work}
\label{sec:related}

We review the related work on inversion attacks, and touch upon
collision attacks as appropriate. Note that for inversion attacks,
every additional round of a hash function inverted is considered
significant improvement over previous work.

\textbf{SAT-based Constructive Methods}.  Since 2005, several hash
functions have been shown to be prone to collision
attacks~\cite{wang2005break,wang2005efficient}. In 2006, Mironov
et.al~\cite{mironov2006applications} automated parts of collision
attack of Wang et al.~\cite{wang2005break} using SAT solvers.  Not
many of the collision attack methods use SAT solvers, as the collision
finding problem is studied very well and most cryptanalysts use direct
implementation of mathematical analyses (e.g., differential
cryptanalysis) for this problem.

Given that the focus of our paper is on inversion attacks, we provide
a thorough overview of the related work for the same.
In 2007, De et al.~\cite{de2007inversion} used a SAT solver for
an inversion attack on MD4 and enhanced number of steps inverted,
up to 2 rounds and 7 steps, by encoding Dobbertin's attack model
\cite{dobbertin1996cryptanalysis} into SAT constraints. In 2008,
Srebrny et al.~\cite{srebrny2008sat} formulated inversion of SHA-1 as
a SAT instance and could solve for restricted message size up to 22
steps.  Several later works could solve up to 23 rounds of SHA-1
\cite{nossum2012sat,legendre2014logical,legendre2012encoding}.
Lafitte et al.~\cite{lafitte2014applications} presented a generic way
to encode basic cryptographic operations which was an improvement over
operator overloading model in \cite{jovanovic2005logical} and used it
in a preimage test on MD4 and finding weak keys in IDEA and MESH
ciphers, although other than finding weak keys, preimage results were
not better than the best previously published attack.  In 2013,
Morawiecki et al.~\cite{morawiecki2013sat} used the idea of
minimization of SAT instances generated via analysis of cryptographic
primitives. They applied their tool CryptLogVer on some hash functions
like SHA-1 and Keccak to analyze their preimage resistance. Although
they did not increase the number of inverted steps, they showed
improvement in solving time. Nossum~\cite{nossum2013instance}, also
presents an encoding for preimage attack of SHA-1 which targets the
5-operand addition operation that is performed in each step of SHA-1
and the generated instances have fewer variables than the work of
Srebrny et al.  \cite{srebrny2008sat} and Morwiecki et
al. \cite{morawiecki2013sat}, and in general are easier to solve by
modern SAT solvers. It is used to generate SHA-1 instances of SAT competition
~\cite{nossum2013instance}.
The work presented in this paper is using Nossum's
instance generator with small tweaks.
All of the mentioned techniques use SAT solvers
as a black box tool. By contrast our design leverages CEGAR and
modifies the restart policies of SAT solvers. Additionally, our method
is faster for finding preimages than previous work.

\textbf{Non SAT Solver based Constructive Attacks}. These methods are
almost exclusively aimed at collision attacks. In 2006, De Canni\`ere
et al.~\cite{de2006finding} built a non SAT solver based tool for
SHA-1 collision attack leveraging the breakthrough of Wang
et al.~\cite{wang2005efficient}. In 2011, Mendel
et al.~\cite{mendel2011finding} extended it for SHA-256 which was
further improved in their work in 2013~\cite{mendel2013improving}.
Eichlseder et al.~\cite{eichlseder2014branching} improved upon the
branching heuristics of this tool and applied it to SHA-512.  Recently
Stevens et al.~\cite{stevens2015freestart} presented a parallelized
search implementation and found free-start collision on full SHA-1.

\textbf{Non-constructive Theoretical Bounds}. Here we review known
theoretical bounds on preimage attack on various hash functions. One
of the first preimage results on SHA-1 was achieved by using
techniques like reversing the inversion problem and mathematical
structures like $P^3$ graphs~\cite{de2008preimages}, which could
invert 34 and 44 steps with complexity of $2^{80}$ and $2^{157}$
respectively.  Aoki and Sasaki~\cite{aoki2009preimage} used
meet-in-the-middle to attack SHA-1 (and also MD4 and MD5) and improved
the number of steps to 48 with the solving complexity of
$2^{159.3}$. Knellwolf et al.~\cite{knellwolf2012new} improved it in
2012 by providing a differential formulation of MITM model and raised
the bar up to 57 steps. Espiatu et al.~\cite{espitau2015higher}
extended this work further to higher order differentials for preimage
attack on SHA-1 and BLAKE2 and went up to 62 steps for
SHA-1. Mathematical structure like
Biclique~\cite{khovratovich2012bicliques}, allowed extending coverage
of MITM over larger number of steps.

\textbf{Adaptive Restarts}. Armin Biere proposed monitoring variable
assignment flips in PicoSAT, and delayed restarts when the weighted
average of flips is below a predetermined
threshold~\cite{biere2008adaptive}.  Audemard and Simon proposed
monitoring the LBD of learnt clauses, and a restart is triggered if
the short term LBDs exceeds the long term LBDs by a constant
factor~\cite{audemard2012refining}. Haim and Walsh used machine
learning to train a classifier to select from a portfolio of restart
policies~\cite{haim2009restart}.  Gagliolo and Schmidhuber used
bandits to select between luby and uniform restart
heuristic~\cite{gagliolo2007learning}.

%% file: sec.conclusion.tex
\section{Conclusion and Future Work}
\label{sec:conclusion}
We presented a tool called \toolname for preimage attack on SHA-1 hash
functions, which uses CEGAR and adaptive restart techniques.  Our tool
is faster than other automated search tools in the literature for the
constructive preimage attack.  Our results show that SAT solvers and
SAT-based techniques are a promising approach for handling the
laborious parts of cryptanalysis, and identifying weaknesses in hash
function designs. This design can be extended to work on other hash
functions like SHA-2 family.